\documentclass[showpacs,aps,graphicx,twocolumn]{revtex4}
\usepackage{graphicx}
\begin{document}

\title{Generalized entanglement distillation}

\author{Yu-Bo Sheng,$^{1}$\footnote{Email address:
shengyb@njupt.edu.cn} Lan Zhou,$^{2}$
 }
\address{
$^1$Institute of Signal Processing  Transmission, Nanjing
University of Posts and Telecommunications, Nanjing, 210003,  China\\
 $^2$College of Mathematics \& Physics, Nanjing University of Posts and Telecommunications, Nanjing,
210003, China\\}

\begin{abstract}
We present a way for entanglement distillation for genuine mixed state.
Different from the conventional mixed state in entanglement purification protocol, each components of the mixed state in our protocol is a less-entangled state, while it is always a maximally entangled state. With the help of the  weak cross-Kerr nonlinearity, this entanglement distillation protocol does not
require the sophisticated single-photon detectors. Moreover, the distilled high quality entangled state can be retained to perform
the further distillation.  It makes more convenient in practical applications.
\end{abstract}
 \maketitle
\section{Introduction}
Entanglement plays an important role in quantum information processing \cite{book,rmp}. Quantum teleportation \cite{teleportation}, quantum dense coding \cite{densecoding1}, quantum
key distribution \cite{Ekert91}, and some other protocols \cite{QSS2,QSDC2} all require the entanglement to set up the maximally entanglement channel.
On the other hand, to complete the quantum computation, they should also create the entanglement \cite{one-way}.
 Unfortunately, during the distribution and storage  of the entanglement, it always
suffers from the environment noise. The environment noise will make the entanglement degrade.
The degraded entanglement will make the quantum communication insecure. It  will also make the
quantum computation cause error.

Generally, the maximally entangled state will degrade to a mixed state.
In an optical system, the maximally entangled state such as the Bell state
$|\phi^{+}\rangle=\frac{1}{\sqrt{2}}(|H\rangle|H\rangle+|V\rangle|V\rangle)$ will become the mixed state as
$\rho_{0}=F|\phi^{+}\rangle\langle\phi^{+}|+(1-F)|\psi^{+}\rangle\langle\psi^{+}|$ \cite{Pan1}.  Here $|\psi^{+}\rangle=\frac{1}{\sqrt{2}}(|H\rangle|V\rangle+|V\rangle|H\rangle)$.
$|H\rangle$ is the horizonal polarization of the photon and $|V\rangle$ is the vertical polarization

of the photon. On the other hand, the maximally entangled state $|\phi^{+}\rangle$ also can degrade to a pure less-entangled state
$|\Phi^{+}\rangle=\gamma|H\rangle|H\rangle+\delta|V\rangle|V\rangle$ \cite{zhao1,Yamamoto1}. Here $|\gamma|^{2}+|\delta|^{2}=1$.
Distilling the high fidelity of  the mixed states from the low quality of the mixed states is the entanglement purification \cite{C.H.Bennett1}. There are a lot
of excellent works focused on the entanglement purification, such as the entanglement purification protocols based on the controlled-Not gate \cite{C.H.Bennett1,D. Deutsch,M. Murao} , linear optics\cite{Pan1,Simon,Pan2,sangouard,sangouard2,shengpra3,lixhepp,dengonestep1,dengonestep2},
nonlinear optics \cite{shengpra1,shengpra2},  and so on \cite{wangc2,wangc4,ren2,shengyb}. The approach  of distilling the pure maximally entangled states from the less-entangled states is entanglement concentration. There are also many excellent works for entanglement concentration, such as the entanglement concentration based on the collective measurement \cite{C.H.Bennett2},
unitary operation \cite{swapping1,swapping2}, linear optics \cite{zhao1,zhao2,Yamamoto1,Yamamoto2,wangxb} and so on \cite{shengpra4,shengsinglephotonconcentration,dengsingle,shengwstateconcentration,ren1}.

Actually,   the decoherence models described above are both existed in a practical entanglement distribution. Unfortunately,
in the previous works, people deal with these problem independently. They either focused on the entanglement purification nor entanglement concentration. Therefore, they can only partially solve the problem of decoherence. In this paper, we will describe a general distillation mode for decoherence. Suppose that
the pure maximally entangled state $|\phi^{+}\rangle$ will both degrade to the mixed state and the less-entangled state, and the decoherence model can be described as
\begin{eqnarray}
\rho=F|\Phi^{+}\rangle\langle\Phi^{+}|+(1-F)|\Psi^{+}\rangle\langle\Psi^{+}|.\label{general0}
\end{eqnarray}

Here we denote
\begin{eqnarray}
|\Phi^{+}\rangle=\gamma|H\rangle|H\rangle+\delta|V\rangle|V\rangle,\label{less1}
\end{eqnarray}
and
\begin{eqnarray}
|\Psi^{+}\rangle=\gamma|H\rangle|V\rangle+\delta|V\rangle|H\rangle.\label{less2}
\end{eqnarray}
From Eq.(\ref{general0}), it is shown that the initial state $|\phi^{+}\rangle$ becomes a mixed state $\rho$, while in each part of the
mixed state, it is still a less-entangled state, say $|\Phi^{+}\rangle$ or $|\Psi^{+}\rangle$. Therefor, in order to distill  such
mixed state $\rho$, we not only need to improve the fidelity of the mixed state $F$, but also concentrate the less-entangled state $|\Phi^{+}\rangle$ and
$|\Psi^{+}\rangle$ to the maximally entangled state. In this paper, we will describe the approach that we can distill such mixed state effectively.
After performing the protocol, we can obtain the high fidelity mixed state with each components being the maximally entangled state. That is, our protocol
can both realize the entanglement purification and entanglement concentration in one step.

This paper is organized as follows: In Sec. II, we will explain our protocol with the correcting of the bit-flip error.
In Sec. III, we will describe the distillation of the phase-flip error. Interestingly, both steps can be repeated to obtain a high success probability.
In Sec.IV, we will extend our protocol to the case of multi-partite entangled systems. In Sec. V, we will make a discussion and conclusion.

\section{Bit-flip error distillation}
Before we start to explain our protocol. It is necessary to introduce the cross-Kerr nonlinearity, which is the key element in
our protocol. From Fig. 1, the Hamiltonian of the cross-Kerr nonlinearity can be written as \cite{QND1,QND2}
\begin{eqnarray}
H_{ck}=\hbar\chi \hat{n}_{a}\hat{n}_{b}.
\end{eqnarray}
If we consider a single photon with $|H\rangle$ polarization in $a1$ spatial mode. This photon combined with the coherent state $|H\rangle|\alpha\rangle$ will evolve to
$|H\rangle|\alpha e^{-i\theta}\rangle$. On the other hand, the  $|V\rangle$ polarization in  $a1$ spatial mode combined with the coherent state $|V\rangle|\alpha\rangle$ will evolve to $|V\rangle|\alpha e^{i\theta}\rangle$. Therefore, by measuring the phase shift of the coherent state, we can judge the single photon number. In this way, we do not need to detect the single photon directly. It is so called the quantum nondemolition (QND) measurement, which has been widely used in quantum information processing \cite{lin1,he1,he2,he3,qi}.

From Fig. 2, suppose that Alice and Bob share two pairs of the mixed states in the spatial modes $a1$, $b1$ and $a2$, $b2$, respectively. The mixed state $\rho$
can be described in Eq. (\ref{general0}).
Before the two pairs passing through the QND, they first perform a bit-flip operation on the second pair in $a2$ and $b2$ modes.
The state  $|\Phi^{+}\rangle$ will become
\begin{eqnarray}
|\Phi^{+}_{1}\rangle=\gamma|V\rangle_{a2}|V\rangle_{b2}+\delta|H\rangle_{a2}|H\rangle_{b2},
\end{eqnarray}
and the $|\Psi^{+}\rangle$ will become
\begin{eqnarray}
|\Psi^{+}_{1}\rangle=\gamma|V\rangle_{a2}|H\rangle_{b2}+\delta|H\rangle_{a2}|V\rangle_{b2}.
\end{eqnarray}
The two photon pairs can be described as follows. With the probability of
$F^{2}$, it is in the state $|\Phi^{+}\rangle_{a1b1}|\Phi^{+}_{1}\rangle_{a2b2}$. With the equal probability of
$F(1-F)$, they are in the state $|\Phi^{+}\rangle_{a1b1}|\Psi^{+}_{1}\rangle_{a2b2}$ and $|\Psi^{+}\rangle_{a1b1}|\Phi^{+}_{1}\rangle_{a2b2}$.
With the probability of $(1-F)^{2}$, it is in the state $|\Psi^{+}\rangle_{a1b1}|\Psi^{+}_{1}\rangle_{a2b2}$.
The items  $|\Phi^{+}\rangle_{a1b1}|\Phi^{+}_{1}\rangle_{a2b2}$ combined with the two coherent states can be described as
\begin{eqnarray}
&&|\Phi^{+}\rangle_{a1b1}|\Phi^{+}_{1}\rangle_{a2b2}|\alpha\rangle_{A}|\alpha\rangle_{B}\nonumber\\
&=&(\gamma|H\rangle_{a1}|H\rangle_{b1}+\delta|V\rangle_{a1}|V\rangle_{b1})\nonumber\\
&&(\gamma|V\rangle_{a2}|V\rangle_{b2}+\delta|H\rangle_{a2}|H\rangle_{b_{2}})|\alpha\rangle_{A}|\alpha\rangle_{B}\nonumber\\
&=&[\gamma^{2}|H\rangle_{a1}|V\rangle_{a2}|H\rangle_{b1}|V\rangle_{b2}+\delta^{2}|V\rangle_{a1}|H\rangle_{a2}|V\rangle_{b1}|H\rangle_{b2}\nonumber\\
&+&\gamma\delta(|H\rangle_{a1}|H\rangle_{a2}|H\rangle_{b1}|H\rangle_{b2}\nonumber\\
&+&|V\rangle_{a1}|V\rangle_{a2}|V\rangle_{b1}|V\rangle_{b2})]|\alpha\rangle_{A}|\alpha\rangle_{B}\nonumber\\
&\rightarrow&\gamma^{2}|H\rangle_{a3}|V\rangle_{a4}|H\rangle_{b3}|V\rangle_{b4}|e^{-2i\theta}\rangle_{A}|e^{-2i\theta}\rangle_{B}\nonumber\\
&+&\delta^{2}|V\rangle_{a3}|H\rangle_{a4}|V\rangle_{b3}|H\rangle_{b4}|e^{2i\theta}\rangle_{A}|e^{2i\theta}\rangle_{B}\nonumber\\
&+&\gamma\delta(|H\rangle_{a3}|H\rangle_{a4}|H\rangle_{b3}|H\rangle_{b4}\nonumber\\
&+&|V\rangle_{a3}|V\rangle_{a4}|V\rangle_{b3}|V\rangle_{b4})|\alpha\rangle_{A}|\alpha\rangle_{B}.\label{evolve1}
\end{eqnarray}
 \begin{figure}[!h]
\begin{center}
\includegraphics[width=8cm,angle=0]{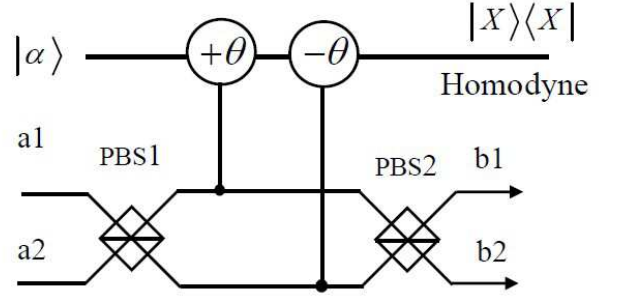}
\caption{Schematic of the quantum nondemolition (QND) measurement with cross-Kerr nonlinearity \cite{shengpra1,qi}. The PBS represent the polarization beam splitter. It can transmit the $|H\rangle$ polarization photon and reflect the $|V\rangle$ polarization photon. The main function of this QND is to make the parity-check measurement, that is to distinguish the states $|H\rangle|H\rangle$, $|V\rangle|V\rangle$ from $|H\rangle|V\rangle$ and $|V\rangle|H\rangle$.}
\end{center}
\end{figure}
With the probability of $(1-F)^{2}$, the states $|\Phi^{+}\rangle_{a1b1}|\Phi^{+}_{1}\rangle_{a2b2}$ combined with the two coherent states
can be written as
\begin{eqnarray}
&&|\Psi^{+}\rangle_{a1b1}|\Psi^{+}_{1}\rangle_{a2b2}|\alpha\rangle_{A}|\alpha\rangle_{B}\nonumber\\
&=&(\gamma|H\rangle_{a1}|V\rangle_{b1}+\delta|V\rangle_{a1}|H\rangle_{b1})\nonumber\\
&&[(\gamma|V\rangle_{a2}|H\rangle_{b2}+\delta|H\rangle_{a2}|V\rangle_{b2})|\alpha\rangle_{A}|\alpha\rangle_{B}\nonumber\\
&=&\gamma^{2}|H\rangle_{a1}|V\rangle_{a2}|V\rangle_{b1}|H\rangle_{b2}+\gamma\delta(|H\rangle_{a1}|H\rangle_{a2}|V\rangle_{b1}|V\rangle_{b2}\nonumber\\
&=&|V\rangle_{a1}|V\rangle_{a2}|H\rangle_{b1}|H\rangle_{b2}\nonumber\\
&+&\delta^{2}|H\rangle_{a2}|V\rangle_{a1}|H\rangle_{b1}|V\rangle_{b2}]|\alpha\rangle_{A}|\alpha\rangle_{B}\nonumber\\
&\rightarrow&\gamma^{2}|H\rangle_{a3}|V\rangle_{a4}|V\rangle_{b3}|H\rangle_{b4}|\alpha e^{-i2\theta}\rangle_{A}|\alpha  e^{i2\theta}\rangle_{B}\nonumber\\
&+&\gamma\delta(|H\rangle_{a3}|H\rangle_{a4}|V\rangle_{b3}|V\rangle_{b4}|\alpha \rangle_{A}|\alpha\rangle_{B}\nonumber\\
&+&|V\rangle_{a3}|V\rangle_{a4}|H\rangle_{b3}|H\rangle_{b4}|\alpha \rangle_{A}|\alpha\rangle_{B}\nonumber\\
&+&\delta^{2}|H\rangle_{a3}|V\rangle_{a4}|H\rangle_{b3}|V\rangle_{b4}|\alpha e^{-i2\theta}\rangle_{A}|\alpha e^{-i2\theta}\rangle_{B}.\label{evolve2}
\end{eqnarray}

In our protocol, the basic principle of distillation is to select the case that both the coherent states pick up no phase shift. In this way, Eq. (\ref{evolve1}) will collapse
to
\begin{eqnarray}
|\phi_{1}\rangle&=&\frac{1}{\sqrt{2}}(|H\rangle_{a1}|H\rangle_{a2}|H\rangle_{b1}|H\rangle_{b2}\nonumber\\
&+&|V\rangle_{a1}|V\rangle_{a2}|V\rangle_{b1}|V\rangle_{b2})).
\end{eqnarray}
The success probability is $2|\gamma\delta|^{2}F^{2}$.
On the other hand, Eq. (\ref{evolve2}) will become
\begin{eqnarray}
|\phi'_{1}\rangle&=&\frac{1}{\sqrt{2}}(|H\rangle_{a3}|H\rangle_{a4}|V\rangle_{b3}|V\rangle_{b4}\nonumber\\
&+&|V\rangle_{a1}|V\rangle_{a2}|H\rangle_{b1}|H\rangle_{b2}).
\end{eqnarray}
The success probability is $2|\gamma\delta|^{2}(1-F)^{2}$. The cross-combination $|\Phi^{+}\rangle_{a1b1}|\Psi^{+}_{1}\rangle_{a2b2}$ and $|\Psi^{+}\rangle_{a1b1}|\Phi^{+}_{1}\rangle_{a2b2}$ never lead both coherent states pick up the same phase shift, and they can be eliminated automatically.
For example,
 with the probability of $F(1-F)$, states $|\Phi^{+}\rangle_{a1b1}|\Psi^{+}_{1}\rangle_{a2b2}$ combined with the two coherent states
can be described as
\begin{eqnarray}
&&|\Phi^{+}\rangle_{a1b1}|\Psi^{+}_{1}\rangle_{a2b2}|\alpha\rangle_{A}|\alpha\rangle_{B}\nonumber\\
&=&(\gamma|H\rangle_{a1}|H\rangle_{b1}+\delta|V\rangle_{a1}|V\rangle_{b1})\nonumber\\
&&(\gamma|V\rangle_{a2}|H\rangle_{b2}+\delta|H\rangle_{a2}|V\rangle_{b_{2}})|\alpha\rangle_{A}|\alpha\rangle_{B}\nonumber\\
&=&[\gamma^{2}|H\rangle_{a1}|V\rangle_{a2}|H\rangle_{b1}|H\rangle_{b2}+\gamma\delta(|H\rangle_{a1}|H\rangle_{a2}|H\rangle_{b1}|V\rangle_{b_{2}})\nonumber\\
&+&|V\rangle_{a1}|V\rangle_{a2}|V\rangle_{b1}|H\rangle_{b2}\nonumber\\
&+&\delta^{2}|V\rangle_{a1}|H\rangle_{a2}|V\rangle_{b1}|V\rangle_{b_{2}}]|\alpha\rangle_{A}|\alpha\rangle_{B}\nonumber\\
&\rightarrow&\gamma^{2}|H\rangle_{a3}|V\rangle_{a4}|H\rangle_{b3}|H\rangle_{b4}|\alpha e^{-i2\theta}\rangle_{A}|\alpha\rangle_{B}\nonumber\\
&+&\gamma\delta|H\rangle_{a3}|H\rangle_{a4}|H\rangle_{b3}|V\rangle_{b4}|\alpha\rangle_{A}|\alpha e^{-i2\theta}\rangle_{B}\nonumber\\
&+&\gamma\delta|V\rangle_{a3}|V\rangle_{a4}|V\rangle_{b3}|H\rangle_{b4}|\alpha\rangle_{A}|\alpha e^{i2\theta}\rangle_{B}\nonumber\\
&+&\delta^{2}|V\rangle_{a3}|H\rangle_{a4}|V\rangle_{b3}|V\rangle_{b4}|\alpha e^{i2\theta}\rangle_{A}|\alpha\rangle_{B}.\label{evolve3}
\end{eqnarray}
From Eq. (\ref{evolve3}), the $|\alpha\rangle_{A}$ will pick up $-2\theta$ phase shift, while $|\alpha\rangle_{B}$ picks up no phase shift, or $|\alpha\rangle_{A}$ will pick up no phase shift, while $|\alpha\rangle_{B}$ picks up  $-2\theta$ phase shift.

Finally, Alice and Bob
both measure the photons in $a4$ and $b4$ spatial modes in $|\pm\rangle=\frac{1}{\sqrt{2}}(|H\rangle+|V\rangle)$.
If the measurement are the same, say $|+\rangle|+\rangle$ or $|-\rangle|-\rangle$, they will obtain the Bell
 state $|\phi^{+}\rangle$ with the fidelity
\begin{eqnarray}
F'=\frac{F^{2}}{F^{2}+(1-F)^{2}}.
\end{eqnarray}
The new mixed state can be written as
\begin{eqnarray}
\rho'=F'|\phi^{+}\rangle\langle\phi^{+}|+(1-F')|\psi^{+}\rangle\langle\psi^{+}|.\label{newmixed}
\end{eqnarray}
On the other hand, if the measurement results are different, say $|+\rangle|-\rangle$ or $|+\rangle|-\rangle$, they will
obtain $|\phi^{-}\rangle=\frac{1}{\sqrt{2}}(|H\rangle|H\rangle-|V\rangle|V\rangle)$, with the same fidelity $F'$. If they obtain $|\phi^{-}\rangle$,
one of the parties say Alice or Bob, only need to perform a Phase-flip operation to transform the $|\phi^{-}\rangle$ to $|\phi^{+}\rangle$.
Therefore, by selecting the cases that both coherent states pick up the same phase shift with 0, they can ultimately increase the fidelity. The total success
probability is $P_{1}=2|\gamma\delta|^{2}[F^{2}+(1-F)^{2}]$.

In above section, we have briefly described the principle of distillation. That is to select the cases that both the coherent states pick up
no phase shift and discard the other cases. Actually, if both coherent states  pick up the phase shift with $2\theta$, the remained states can be reused
 to increase the total success probability. Here they should make the $\pm2\theta$ undistinguished.   This measurement can be achieved by choosing the local oscillator phase $\pi/2$ offset from the
 probe phase. It is called an $X$ quadrature measurement \cite{QND1}.
In this way, Eq.(\ref{evolve1}) will become
\begin{eqnarray}
|\phi_{2}\rangle&=&\frac{\gamma^{2}}{\sqrt{|\gamma|^{4}+|\delta|^{4}}}|H\rangle_{a3}|H\rangle_{a4}|H\rangle_{b3}|H\rangle_{b4}\nonumber\\
&+&\frac{\delta^{2}}{\sqrt{|\gamma|^{4}+|\delta|^{4}}}|V\rangle_{a3}|V\rangle_{a4}|V\rangle_{b3}|V\rangle_{b4}.
\end{eqnarray}
On the other hand, Eq. (\ref{evolve2}) will become
\begin{eqnarray}
|\phi'_{2}\rangle&=&\frac{\gamma^{2}}{\sqrt{|\gamma|^{4}+|\delta|^{4}}}|H\rangle_{a3}|V\rangle_{a4}|H\rangle_{b3}|V\rangle_{b4}\nonumber\\
&+&\frac{\delta^{2}}{\sqrt{|\gamma|^{4}+|\delta|^{4}}}|V\rangle_{a3}|H\rangle_{a4}|V\rangle_{b3}|H\rangle_{b4}.
\end{eqnarray}
Finally, by measuring the photons in the spatial modes $a4$ and $b4$ in the $|\pm\rangle$ basis, they will ultimately obtain a new mixed state with
\begin{eqnarray}
\rho_{1}=F'|\Phi^{+}_{1}\rangle\langle\Phi^{+}_{1}|+(1-F')|\Psi^{+}_{1}\rangle\langle\Psi^{+}_{1}|.
\end{eqnarray}
Here
\begin{eqnarray}
|\Phi^{+}_{1}\rangle&=&\frac{\gamma^{2}}{\sqrt{|\gamma|^{4}+|\delta|^{4}}}|H\rangle_{a3}|H\rangle_{b3}\nonumber\\
&+&\frac{\delta^{2}}{\sqrt{|\gamma|^{4}+|\delta|^{4}}}|V\rangle_{a3}|V\rangle_{b3},
\end{eqnarray}
and
\begin{eqnarray}
|\Psi^{+}_{1}\rangle&=&\frac{\gamma^{2}}{\sqrt{|\gamma|^{4}+|\delta|^{4}}}|H\rangle_{a3}|V\rangle_{b3}\nonumber\\
&+&\frac{\delta^{2}}{\sqrt{|\gamma|^{4}+|\delta|^{4}}}|V\rangle_{a3}|H\rangle_{b3}.
\end{eqnarray}
In the next step, Alice prepares a single photon of the form
\begin{eqnarray}
|\varphi_{1}\rangle=\frac{\gamma^{2}}{\sqrt{|\gamma|^{4}+|\delta|^{4}}}|H\rangle_{a2}+\frac{\delta^{2}}{\sqrt{|\gamma|^{4}+|\delta|^{4}}}|V\rangle_{a2}.\label{single1}
\end{eqnarray}
Then she lets his photon in $a2$ spatial modes and the single photon pass through the QND shown in Fig. 1. After two photons passing through the
QND, if the coherent state picks up no phase shift, they will obtain the  state $|\phi^{+}\rangle$ with the probability of
$\frac{2|\gamma\delta|^{4}}{|\gamma|^{4}+|\delta|^{4}}F^{2}$. They also will obtain the  $|\psi^{+}\rangle$ with the probability of
$\frac{2|\gamma\delta|^{4}}{|\gamma|^{4}+|\delta|^{4}}(1-F)^{2}$. In this way, they can get the mixed state $\rho'$ with the success probability of $\frac{2|\gamma\delta|^{4}}{|\gamma|^{4}+|\delta|^{4}}[F^{2}+(1-F)^{2}]$.
Interestingly, if the coherent states pick up the phase shift with $2\theta$, they will obtain another mixed state $\rho_{2}$ with
\begin{eqnarray}
\rho_{2}=F'|\Phi^{+}_{2}\rangle\langle\Phi^{+}_{2}|+(1-F')|\Psi^{+}_{2}\rangle\langle\Psi^{+}_{2}|,
\end{eqnarray}
with
\begin{eqnarray}
|\Phi^{+}_{2}\rangle&=&\frac{\gamma^{4}}{\sqrt{|\gamma|^{8}+|\delta|^{8}}}|H\rangle_{a3}|H\rangle_{b3}\nonumber\\
&+&\frac{\delta^{4}}{\sqrt{|\gamma|^{8}+|\delta|^{8}}}|V\rangle_{a3}|V\rangle_{b3},
\end{eqnarray}
and
\begin{eqnarray}
|\Psi^{+}_{2}\rangle&=&\frac{\gamma^{4}}{\sqrt{|\gamma|^{8}+|\delta|^{8}}}|H\rangle_{a3}|V\rangle_{b3}\nonumber\\
&+&\frac{\delta^{4}}{\sqrt{|\gamma|^{8}+|\delta|^{8}}}|V\rangle_{a3}|H\rangle_{b3}.
\end{eqnarray}
They can also distill the maximally entangled state from $\rho_{2}$ assisted with another single photon of the form
\begin{eqnarray}
|\varphi_{2}\rangle=\frac{\gamma^{4}}{\sqrt{|\gamma|^{8}+|\delta|^{8}}}|H\rangle_{a2}+\frac{\delta^{4}}{\sqrt{|\gamma|^{8}+|\delta|^{8}}}|V\rangle_{a2}.
\end{eqnarray}

\begin{figure}[!h]
\begin{center}
\includegraphics[width=8cm,angle=0]{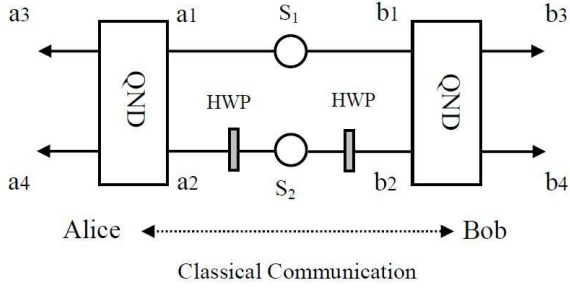}
\caption{The entanglement distillation protocol using QND measurement. HWP represents the half-wave plate, which can transform the $|H\rangle$ to $|V\rangle$, and $|V\rangle$ to $|H\rangle$, respectively.}
\end{center}
\end{figure}
In this way, it can be repeated for $N$ round to increase the total success probability. In the first round,
the success probability $P_{1}$ can be described as
\begin{eqnarray}
P_{1}=\frac{2|\gamma\delta|^{2}}{|\gamma|^{2}+|\delta|^{2}}[F^{2}+(1-F)^{2}].
\end{eqnarray}
We can get $P_{2}$ as
\begin{eqnarray}
P_{2}=\frac{2|\gamma\delta|^{4}}{(|\gamma|^{2}+|\delta|^{2})(|\gamma|^{4}+|\delta|^{4})}[F^{2}+(1-F)^{2}].
\end{eqnarray}
We can also get
\begin{eqnarray}
P_{N}&=&\frac{2|\gamma\delta|^{2^{N}}}{(|\gamma|^{2}+|\delta|^{2})(|\gamma|^{4}+|\delta|^{4})\cdots(|\gamma|^{2^{N}}+|\delta|^{2^{N}})}\nonumber\\
&\times&[F^{2}+(1-F)^{2}].
\end{eqnarray}
The total success probability of obtaining the mixed state $\rho'$ is
\begin{eqnarray}
P_{t}&=&\sum^{N}_{i=1}P_{i}=(\sum^{N}_{i=1}\frac{2|\gamma\delta|^{2^{i}}}{\prod^{N}_{i=1}(|\gamma|^{2^{i}}+|\delta|^{2^{i}})})\nonumber\\
&\times&[F^{2}+(1-F)^{2}].
\end{eqnarray}

\section{Phase-flip error distillation}
In this section, we will describe the distillation of the phase-flip error. In the previous entanglement purification
protocols, the phase-flip error can be conversed to the bit-flip error by Hadamard operation and can be purified in the next round.
However, in this protocol, we cannot treat it like the previous entanglement purification protocols. Suppose that Alice and Bob share the mixed state of the
form
\begin{eqnarray}
\rho_{p}=F|\Phi^{+}\rangle\langle\Phi^{+}|+(1-F)|\Phi^{-}\rangle\langle\Phi^{-}|.\label{general2}
\end{eqnarray}
Here
\begin{eqnarray}
|\Phi^{-}\rangle=\gamma|H\rangle|H\rangle-\delta|V\rangle|V\rangle.
\end{eqnarray}
Alice first prepares a single photon of the form
\begin{eqnarray}
|\varphi_{0}\rangle=\gamma|H\rangle+\delta|V\rangle.
\end{eqnarray}
The state $\rho_{p}$ is in the $a1$ and $b1$ spatial modes and  $|\varphi_{0}\rangle$ is in the $a2$ spatial mode. Alice first let his two photons pass through the QND.
The state $|\Phi^{+}\rangle_{a1b1}$ and $|\varphi_{0}\rangle_{a2}$ combined with the coherent state can be described as
\begin{eqnarray}
&&|\Phi^{+}\rangle_{a1b1}|\varphi_{0}\rangle_{a2}|\alpha\rangle_{A}\nonumber\\
&=&(\gamma|H\rangle_{a1}|H\rangle_{b1}+\delta|V\rangle_{a1}|V\rangle_{b1})(\gamma|H\rangle_{a2}+\delta|V\rangle_{a2})|\alpha\rangle_{A}\nonumber\\
&\rightarrow&(\gamma|H\rangle_{a1}|H\rangle_{b1}+\delta|V\rangle_{a1}|V\rangle_{b1})(\gamma|V\rangle_{a2}+\delta|H\rangle_{a2})|\alpha\rangle_{A}\nonumber\\
&\rightarrow&\gamma^{2}|H\rangle_{a3}|V\rangle_{a4}|H\rangle_{b3}|\alpha e^{i2\theta}\rangle_{A}\nonumber\\
&+&\delta^{2}|V\rangle_{a3}|H\rangle_{a4}|V\rangle_{b3}|\alpha e^{-i2\theta}\rangle_{A}\nonumber\\
&+&\gamma\delta(|H\rangle_{a3}|H\rangle_{a4}|H\rangle_{b3}+|V\rangle_{a3}|V\rangle_{a4}|V\rangle_{b3})|\alpha\rangle_{A}.
\end{eqnarray}

Obviously, if the coherent state picks up no phase shift,
they will obtain the state $|\phi^{+}\rangle$ after measuring the photon in $a3$ spatial modes in $|\pm\rangle$ basis, with the probability of $2|\gamma\delta|^{2}F$. On the other hand, if the initial state is $|\Phi^{-}\rangle$,
they will obtain $|\phi^{-}\rangle$
 with the probability of $2|\gamma\delta|^{2}(1-F)$.
In this way, the mixed state $\rho_{p}$ becomes a new mixed state
\begin{eqnarray}
\rho'_{p}=F|\phi^{+}\rangle\langle\phi^{+}|+(1-F)|\phi^{-}\rangle\langle\phi^{-}|.\label{general3}
\end{eqnarray}
The success probability is $2|\gamma\delta|^{2}$.
Certainly, they can also repeat the protocol to increase the success probability, if the coherent state picks up the phase shift $2\theta$.
In the second round, Alice only needs to prepare another single photon of the form of $|\varphi_{1}\rangle$ as shown in Eq. (\ref{single1}).
Following the same principle, if the coherent state picks up no phase shift, they will obtain $\rho'_{p}$ with the success probability
of $\frac{2|\gamma\delta|^{4}}{|\gamma|^{4}+|\delta|^{4}}$. If they repeat it for $N$ round, the success probability is
\begin{eqnarray}
P_{p}&=&2|\gamma\delta|^{2}+\frac{2|\gamma\delta|^{4}}{|\gamma|^{4}+|\delta|^{4}}+\cdots\nonumber\\
&+&\frac{2|\gamma\delta|^{2^{N}}}{(|\gamma|^{2}+|\delta|^{2})(|\gamma|^{4}+|\delta|^{4})\cdots(|\gamma|^{2^{N}}+|\delta|^{2^{N}})}\nonumber\\
&=&\sum^{N}_{i=1}\frac{2|\gamma\delta|^{2^{i}}}{\prod^{N}_{i=1}(|\gamma|^{2^{i}}+|\delta|^{2^{i}})}.
 \end{eqnarray}
If they obtain the state $\rho'_{p}$, it is the standard mixed state with a phase-flip error like Ref. \cite{Pan1}. In this way, they
both perform the Hadamard  operation to converse the phase-flip error to a bit-flip error. After performing the Hadamard operations on
two photons, the state of $|\phi^{+}\rangle$ does not change, while $|\phi^{-}\rangle$ becomes $|\phi^{+}\rangle$. The form of the transformed  mixed state is
the same as the mixed state shown in Eq. (\ref{newmixed}). Therefore, it can be distilled in a next round.

\section{Distillation of multipartite Greenberg-
Horne-Zeilinger state}
In this section, we will describe the distillation of the multipartite Greenberg-Horne-Zeilinger (GHZ) state.
An $N$-particle GHZ state can be described as
\begin{eqnarray}
|\phi^{+}_{N}\rangle=\frac{1}{\sqrt{2}}(|H\rangle|H\rangle\cdots|H\rangle_{N}+|V\rangle|V\rangle\cdots|V\rangle_{N}).
 \end{eqnarray}
 Suppose that the mixed state can be written as
 \begin{eqnarray}
 \rho_{N}=F|\Phi^{+}_{N}\rangle\langle\Phi^{+}_{N}|+(1-F)|\Psi^{+}_{N}\rangle\langle\Psi^{+}_{N}|.\label{GHZ1}
 \end{eqnarray}

\begin{figure}[!h]
\begin{center}
\includegraphics[width=8cm,angle=0]{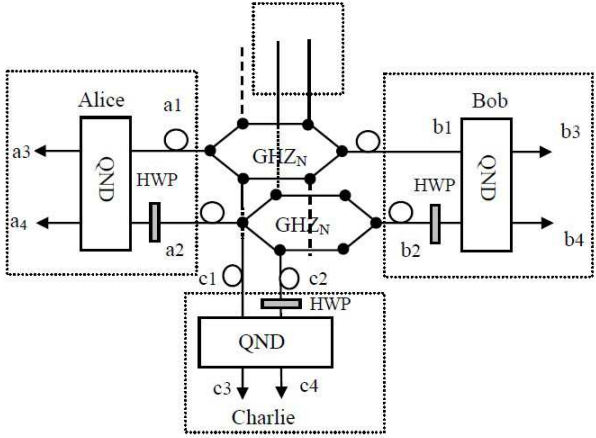}
\caption{The entanglement distillation protocol for multipartite GHZ state with QNDs. The GHZ$_{N}$ represents the $N$-particle GHZ state.}
\end{center}
\end{figure}
Here
 \begin{eqnarray}
 |\Phi^{+}_{N}\rangle=\gamma|H\rangle_{1}|H\rangle_{2}\cdots|H\rangle_{N}+\delta|V\rangle_{1}|V\rangle_{2}\cdots|V\rangle_{N},\label{GHZ2}
 \end{eqnarray}
 and
 \begin{eqnarray}
 |\Psi^{+}_{N}\rangle=\gamma|V\rangle_{1}|H\rangle_{2}\cdots|H\rangle_{N}+\delta|H\rangle_{1}|V\rangle_{2}\cdots|V\rangle_{N}.\label{GHZ3}
 \end{eqnarray}
 The principle of distillation the multipartite GHZ state is similar to the previous description.
 The $N$ photons are distributed to Alice, Bob, Charlie, etc. In each round, they choose two pairs of mixed state $\rho_{N}$.
 The first pair is in the spatial modes $a1$, $b1$, $c1$, and so on. The second pair is in the  $a2$, $b2$, $c2$, and so on.
 They first let the state $ |\Phi^{+}_{N}\rangle$ perform  bit-flip operation on each photon and make the whole state become
 \begin{eqnarray}
 |\Phi^{+}_{N}\rangle'=\gamma|V\rangle_{1}|V\rangle_{2}\cdots|V\rangle_{N}+\delta|H\rangle_{1}|H\rangle_{2}\cdots|H\rangle_{N}.\label{GHZ4}
 \end{eqnarray}
 The state $ |\Psi^{+}_{N}\rangle$ becomes
 \begin{eqnarray}
 |\Psi^{+}_{N}\rangle'=\gamma|H\rangle_{1}|V\rangle_{2}\cdots|V\rangle_{N}+\delta|V\rangle_{1}|H\rangle_{2}\cdots|H\rangle_{N}.\label{GHZ5}
 \end{eqnarray}
 The new mixed state can be written as
  \begin{eqnarray}
 \rho'_{N}=F|\Phi^{+}_{N}\rangle''\langle\Phi^{+}_{N}|+(1-F)|\Psi^{+}_{N}\rangle''\langle\Psi^{+}_{N}|.\label{GHZ6}
 \end{eqnarray}

As shown in Fig. 3, they choose two copies of mixed states in each round.
The whole system can also be regarded as the mixture of four states. With the probability of
$F^{2}$, it is in the state $|\Phi^{+}_{N}\rangle|\Phi^{+}_{N}\rangle'$. With the  same probability of
$F(1-F)$, they are in the states $|\Phi^{+}_{N}\rangle |\Psi^{+}_{N}\rangle'$ and $|\Psi^{+}_{N}\rangle |\Phi^{+}_{N}\rangle'$.
With the probability of $(1-F)^{2}$, it is in the state $ |\Psi^{+}_{N}\rangle |\Psi^{+}_{N}\rangle'$.
The principle of the distillation is similar to the previous one. After the photons passing through the QNDs, they pick up the  case that
all the coherent states have no phase shift. In this way, the cross-combination items can be eliminated automatically. The remained states
are
\begin{eqnarray}
|\phi^{+}_{2N}\rangle=\frac{1}{\sqrt{2}}(|H\rangle_{1}|H\rangle_{2}\cdots|H\rangle_{2N}+|V\rangle_{1}|V\rangle_{2}\cdots|V\rangle_{2N}),\nonumber\\
 \end{eqnarray}
with the probability of $2|\gamma\delta|^{2}F^{2}$, and the state
\begin{eqnarray}
|\psi^{+}_{2N}\rangle&=&\frac{1}{\sqrt{2}}(|V\rangle_{1}|V\rangle_{2}|H\rangle_{3}\cdots|H\rangle_{2N}\nonumber\\
&+&|H\rangle_{1}|H\rangle_{2}|V\rangle_{3}\cdots|V\rangle_{2N}),
 \end{eqnarray}
with the probability of $2|\gamma\delta|^{2}(1-F)^{2}$.
Finally, by measuring the even number of the photons in the basis $|\pm\rangle$, they will ultimately obtain the
new mixed state
\begin{eqnarray}
 \rho''_{N}=F'|\phi^{+}_{N}\rangle\langle\phi^{+}_{N}|+(1-F')|\psi^{+}_{N}\rangle\langle\psi^{+}_{N}|.
  \end{eqnarray}
  Here
  \begin{eqnarray}
  |\psi^{+}_{N}\rangle=\frac{1}{\sqrt{2}}(|V\rangle|H\rangle\cdots|H\rangle_{N}+|H\rangle|V\rangle\cdots|V\rangle_{N}).
    \end{eqnarray}
 \begin{figure}[!h]
\begin{center}
\includegraphics[width=7cm,angle=0]{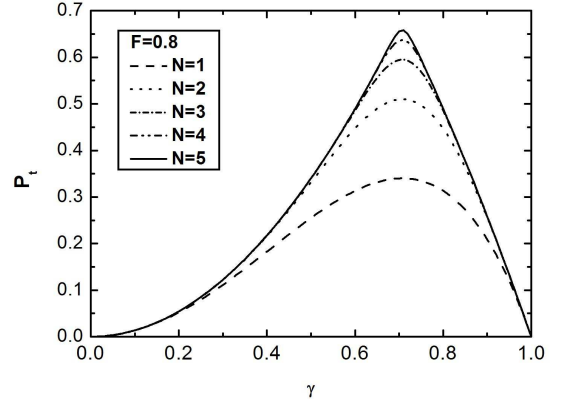}
\caption{The success probability $P_{t}$ is altered with the initial coefficient $\gamma$. Here we let $F=0.8$. $N$ is the iteration number.}
\end{center}
\end{figure}

 On the other hand, if they pick up the same phase shift $2\theta$, and measuring the even number photons in the
 $|\pm\rangle$ basis, they will obtain
 \begin{eqnarray}
|\Phi^{+}_{N}\rangle&=&\frac{\gamma^{2}}{\sqrt{|\gamma|^{4}+|\delta|^{4}}}|H\rangle_{1}|H\rangle_{2}\cdots|H\rangle_{2N}\nonumber\\
&+&\frac{\delta^{2}}{\sqrt{|\gamma|^{4}+|\delta|^{4}}}|V\rangle_{1}|V\rangle_{2}\cdots|V\rangle_{2N}\nonumber\\
\end{eqnarray}
and
\begin{eqnarray}
|\Psi^{+}_{N}\rangle&=&\frac{\gamma^{2}}{\sqrt{|\gamma|^{4}+|\delta|^{4}}}|V\rangle_{1}|H\rangle_{2}|H\rangle_{3}\cdots|H\rangle_{2N}\nonumber\\
&+&\frac{\delta^{2}}{\sqrt{|\gamma|^{4}+|\delta|^{4}}}|H\rangle_{1}|V\rangle_{2}\cdots|V\rangle_{2N}.
\end{eqnarray}
Similarly, they only need to prepare the single photon state of the form as show in Eq. (\ref{single1}), and following the same operation
shown in the previous section, they will obtain the same mixed state $\rho''_{N}$ with the probability of $\frac{2|\gamma\delta|^{4}}{|\gamma|^{4}+|\delta|^{4}}[F^{2}+(1-F)^{2}]$. In this way, it can also be repeated to obtain a high success probability.
Once the bit-flip error can be corrected, the phase flip error can also be corrected in a next round.

\section{discussion and conclusion}
So far, we have fully described our protocol. We first explained the protocol with a bit-flip error. By selecting the cases that both
coherent states picking up no phase shift, they can obtain the higher fidelity of the mixed state. On the other hand, if both the coherent states
pick up the phase shift with $2\theta$, this protocol can be repeated to reach a high success probability. In our protocol, we also show that the phase-flip error can also be well distilled. Different from the bit-flip error correction, it can be achieved in two steps. In the first step, they first prepare a single photon. After the single photon and the mixed state both pass through the QND in Alice's location, they select the case that the coherent state pick up no phase shift. In this way, the original mixed state becomes a standard mixed state in Eq. (\ref{general3}). Subsequently, they perform the Hadamard operation on both photons to convert the phase-flip error to the bit-flip error, which can be distilled in a conventional way.  Interestingly, in the first step, if the coherent state pick up the phase shift $2\theta$, it can also be repeated to obtain a high success probability.

It is interesting to compare this protocol with previous entanglement purification protocols and concentration protocols. In the pioneer work of purification \cite{Pan1},
they  achieved the bit-flip error correction with the post-selection principle. Ref. \cite{zhao1} also described the entanglement concentration protocol. In their protocol, they can obtain the maximally entangled state with the success probability of $2|\alpha\beta|^{2}$. In our protocol, if we let $F=1$, our model
is simplified to the standard entanglement concentration. On the other hand, if we let $\gamma=\delta=\frac{1}{\sqrt{2}}$, this protocol is essentially the entanglement purification model. Therefore, our distillation protocol is more practical and universal than both purification and concentration protocols. Actually, if the pure maximally entangled state degrades to the mixed state shown in Eq. (\ref{general0}), one can also perform the entanglement concentration first to obtain the standard mixed state in Eq. (\ref{general3}) and perform the entanglement purification subsequently. In our protocol, one of the advantage is that it can be completed in one step.
Moreover, this protocol can be repeated to obtain a high success probability.  In Fig. 4, we calculated the success probability altered with the initial coefficient $\gamma$ in the distillation of the bit-flip error.
We also let $F=0.8$.  From Fig. 4, it is shown that the success probability increases greatly if we repeat this protocol. The max success probability is only 0.34 with $N=1$, while it can reach 0.66 with $N=5$.

In our protocol, the cross-Kerr nonlinearity plays an important role in distillation. The success case is that the coherent state picks up no phase shift.
It is essentially to make a parity check. They can distinguish the even parity state $|H\rangle|H\rangle$ and $|V\rangle|V\rangle$ from the odd parity state
 $|H\rangle|V\rangle$ and $|V\rangle|H\rangle$, according to the different phase shifts of the coherent states. In a practical experiment, we should require
 the phase shift of the cross-Kerr nonlinearity to reach a visible value. That is to require the $\alpha\theta>1$. One method is to increase the amplitude of the coherent state. The other method is to amplify the normal cross-Kerr nonlinearity. Recently theoretical  work showed that the giant Kerr nonlinearity can be obtain in a
 multiple quantum-well structure with a four-level, double $\wedge$-type configuration \cite{oe}. On the other hand, the weak measurement can also be used to amplify the cross-Kerr nonlinearity \cite{weak_meaurement}. Moreover, current experiment reported that the "giant" cross-Kerr effect with phase shift of 20 degrees per photon has been observed \cite{gaint}.

 In conclusion, we presented a practical entanglement distillation protocol for general mixed state model. In the general mixed state model, each components
 are still the less-entangled states. Therefore, our protocol can not only improve the fidelity of the mixed state, but also can concentrate the less-entangled state to the maximally entangled state. For the bit-flip correction, the distinct advantage is that such process can be completed in one step.  Moreover, this protocol can reach a high success probability by repeating this protocol, resorting to the QND measurement. This protocol has practical application in the future quantum information processing.

\section*{ACKNOWLEDGEMENTS}
This work is supported by the National Natural Science Foundation of
China under Grant Nos. 11104159 and 11347110, and the Project
Funded by the Priority Academic Program Development of Jiangsu
Higher Education Institutions.

\end{document}